\documentclass[a4paper,12pt]{article}
\usepackage[pctex32]{graphics}
\usepackage{amssymb,amsmath}

\begin{document}

\topmargin 0pt
\oddsidemargin 0mm
\newcommand{\be}{\begin{equation}}
\newcommand{\ee}{\end{equation}}
\newcommand{\ba}{\begin{eqnarray}}
\newcommand{\ea}{\end{eqnarray}}
\newcommand{\fr}{\frac}
\renewcommand{\thefootnote}{\fnsymbol{footnote}}

\begin{titlepage}

\begin{center}
{\Large \bf  Absorption cross section in Lifshitz black hole  }

\vskip .6cm {\large   Taeyoon Moon$^{a}$ \footnote{e-mail
 address: tymoon@sogang.ac.kr} and Yun Soo Myung$^{b}$ \footnote{e-mail address: ysmyung@inje.ac.kr}}
\\
\vspace{10mm} {{${}^{a}${\em Center for Quantum Space-time, Sogang
University, Seoul, 121-742, Korea}
\\
${}^{b}${\em Institute of Basic Science and School of Computer Aided
Science,  Inje University\\ Gimhae 621-749, Korea}}} \vspace{5mm}
\vskip .6cm

\end{center}

\begin{center}

\underline{Abstract}
\end{center}
We derive the absorption cross section of a minimally coupled scalar
in the  Lifshitz black hole  obtained from the new massive gravity.
The  absorption cross section reduces to the horizon area in the low
energy and massless limits of scalar propagation, indicating that
the Lifshitz black hole also satisfies  the universality of
low-energy absorption cross section for black holes. \vspace{5mm}


\thispagestyle{empty}
\end{titlepage}
\renewcommand{\thefootnote}{\arabic{footnote}}
\setcounter{footnote}{0}
\newpage
\section{Introduction}
The Lifshitz-type black
holes~\cite{L-0,CFT-4,L-1,AL-3,L-2,L-4,L-3,L-5,CFT-1,CFT-2,CFT-3,Mann,bm,tay,MTr,BBP,AL-2}
have received considerable attentions since these may provide a
model of generalizing AdS/CFT correspondence to non-relativistic
condensed matter physics [the Lif/CFT correspondence]. Although
their asymptotic spacetimes  are  known to be Lifshitz, the whole
properties of these black holes are not yet explored completely.

Especially, we wish to focus on the ($z=3$) Lifshitz black
hole~\cite{z3} derived from the new massive gravity~\cite{bht}
because it may be considered as a toy model for Lifshitz black
holes.

If a black hole is found,  a thermodynamic study  is important to
understand the black hole because heat capacity and free
energy determine the thermodynamic  stability of the  black
hole.  Recently, there was a progress on computation of mass and
related thermodynamic quantities by using the ADT
method~\cite{DS-1,DS-2} and Euclidean action approach in three
dimensions~\cite{GTT}. There was a discrepancy in mass  between
${\cal M}=\frac{7r_+^4}{8G \ell^4}$ obtained from the ADT
method~\cite{DS-1} and ${\cal M}=\frac{r_+^4}{4G \ell^4}$ from
other approaches~\cite{GTT,HT,MKP}.    However,  it turned out that
the  Lifshitz black hole is thermodynamically stable since its
heat capacity is positive and free energy is negative.  The
possibility of phase transition between Lifshitz black hole and
thermal Lifshitz has been discussed  by introducing  on-shell and
off-shell free energies~\cite{myung}.

On the other hand, quasinormal modes  of a black hole contain
important information on the black hole.  Its complex quasinormal
frequency is given by $\omega=\omega_R-i\omega_I$  whose real part
represents the oscillation and imaginary part denotes the rate at
which this oscillation is damped, because of the very nature of
black hole horizon.   In addition,  the condition of $\omega_I>0$ is
consistent with the stability condition of the black hole.
Quasinormal frequencies (QNFs) were obtained from a perturbed scalar
propagation  by imposing the boundary condition: ingoing mode near
horizon and Dirichlet boundary condition at infinity.  Importantly,
QNFs of Lifshitz black hole are purely imaginary~\cite{COP,MM}, showing  that such
perturbation has no oscillation stage.  This feature may be related
to the solitonic nature on the boundary field theory which implies
that its equilibrium is  comparatively stable and thus, it is  difficult
to take the theory out of equilibrium~\cite{Abdalla:2011fd}.    Consequently, this
indicates  that {\it the Lifshitz black hole is stable against an external
perturbation, which is closely related to its thermodynamic
stability.}

At this stage, one has to ask how the Lifshitz black hole is
different from a three-dimensional known black hole of  the
non-rotating BTZ black hole~\cite{BTZ1,BTZ2}. One difference is that
QNFs of Lifshitz black hole are purely imaginary, whereas those of
BTZ black hole are complex. Their thermodynamic property is the
nearly same to each other: positive heat capacity and negative free
energy~\cite{myungbtz} even though they have different asymptotes.
One remaining  thing to explore is to compute the absorption cross
section (=greybody factor) because it provides a litmus to test
whether or not the Lifshitz black hole possesses  the universal
property of black holes~\cite{DGM}.

In this work, we obtain the absorption cross section  by
investigating a minimally coupled scalar propagating in the Lifshitz
black hole background. Importantly, we will show that the absorption
cross section reduces to the horizon area in the low energy and
massless limits of $s$-wave  propagation, indicating that the
Lifshitz black hole also satisfies the universality of low energy
absorption cross section for any black holes.

\section{Lifshitz black hole from new massive gravity}

We start with the new massive gravity~\cite{bht} composed of the
Einstein-Hilbert action with a cosmological constant $\Lambda$ and
higher-order curvature terms given by
\begin{eqnarray}
\label{NMGAct}
 S_{\rm NMG} &=&-\Big[ S_{\rm EH}+S_{\rm HC}\Big], \\
\label{NMGAct2} S_{\rm EH} &=& \frac{1}{16\pi G} \int d^3x \sqrt{-g}~ (R-2\Lambda),\\
\label{NMGAct3} S_{\rm HC} &=& -\frac{1}{16\pi G\tilde{m}^2} \int
d^3x
            \sqrt{-g}~\left(R_{\mu\nu}R^{\mu\nu}-\frac{3}{8}R^2\right),
\end{eqnarray}
where $G$ is a three-dimensional Newton constant and $\tilde{m}^2$ a
parameter with mass dimension 2.  We would like  to mention that to
avoid negative mass and entropy, it is necessary to take ``$-$" sign
in the front of $[ S_{\rm EH}+S_{\rm HC}]$.   The field equation is
given by \be R_{\mu\nu}-\frac{1}{2}g_{\mu\nu}R+\Lambda
g_{\mu\nu}-\frac{1}{2\tilde{m}^2}K_{\mu\nu}=0,\ee where
\begin{eqnarray}
  K_{\mu\nu}&=&2\square R_{\mu\nu}-\frac{1}{2}\nabla_\mu \nabla_\nu R-\frac{1}{2}\square{R}g_{\mu\nu}\nonumber\\
        &+&4R_{\mu\nu\rho\sigma}R^{\rho\sigma} -\frac{3}{2}RR_{\mu\nu}-R_{\rho\sigma}R^{\rho\sigma}g_{\mu\nu}
         +\frac{3}{8}R^2g_{\mu\nu}.
\end{eqnarray}
In order to obtain the $z=3$  Lifshitz black hole solution, we have to choose
 $\tilde{m}^2=-\frac{1}{2\ell^2}$ and
$\Lambda=-\frac{13}{2\ell^2}$ with  $\ell$  the curvature radius of
Lifshitz  spacetimes.  Explicitly, we find the $z=3$ Lifshitz black
hole solution~\cite{z3} as
 \be
\label{3dmetric}
  ds^2_{\rm Lif}=g_{\mu\nu}dx^\mu dx^\nu=-\left(\frac{r^2}{\ell^2}\right)^z\left(1-\frac{M\ell^2}{r^2}\right)dt^2
   +\frac{dr^2}{\left(\frac{r^2}{\ell^2}-M\right)}+r^2d\phi^2,
\end{equation}
where $M$ is an integration constant related to the mass of black
hole. The event horizon is located at $r=r_+=\ell \sqrt{M}$. The
$z=1$ case corresponds to the BTZ black hole. The line element
(\ref{3dmetric}) is invariant under the anisotropic scaling of
\begin{equation}
t\to \lambda^zt,~~\phi \to \lambda \phi,~~ r\to \frac{r}{\lambda}
\end{equation}
with $M\to M/\lambda^2$. For $z=1$ BTZ black hole, the ADM mass is
determined to be $M=\frac{r_+^2}{\ell^2}$, while for $z=3$ Lifshitz
black hole, the ADM mass is proportional to  $M^2$. Its
$z$-dependent curvature is given by \be R_z=\frac{2}{\ell^2
r^2}\Big[\ell^2M-r^2-(2\ell^2M+r^2)z+(\ell^2M-r^2)z^2\Big], \ee
which yields \be
R_{z=1}=-\frac{6}{\ell^2},~~~R_{z=3}=-\frac{26}{\ell^2}+\frac{8M}{r^2}.
\ee

\section{Scalar propagation in Lifshitz spacetimes}
In order to find the absorbtion cross section, we first consider a
minimally coupled scalar described by the Klein-Gordon equation
 \be \Big[\bar{\square}_{\rm Lif}-m^2\Big] \varphi=0 \ee in the
background of Lifshitz black hole spacetimes (\ref{3dmetric}) which
yields
\begin{eqnarray}\label{omain}
(r^2-r_+^2)\rho''(r)+\frac{5r^2-3r_+^2}{r}\rho'(r)+
\frac{\ell^8\omega^2-r^2(\ell^2k^2+m^2\ell^2r^2)(r^2-r_+^2)}{r^4(r^2-r_+^2)}
\rho(r)=0
\end{eqnarray}
for the ansatz $\varphi=\rho(r)e^{-i\omega t + ik \phi}$. Now we
consider a tortoise coordinate $r^{*}$ as
\begin{eqnarray}\label{toto}
r^{*}=\frac{\ell^4}{r_+^3}\left[ \frac{r_+}{r}-{\rm
arccoth}\left(\frac{r}{r_+}\right)\right],
\end{eqnarray}
which is defined by $dr^{*}=dr/f(r)$ with
$f(r)=r^2(-M+r^2/\ell^2)/\ell^2$. Then $r\in[r_+,\infty]$ is mapped
into $r^*\in[-\infty,0]$. Introducing a new field
$\Phi$($=\sqrt{r}\rho(r)$) together with $r^*$, Eq.(\ref{omain}) can
be written as the Schr\"{o}dinger-type equation
\begin{eqnarray}\label{sch}
\frac{d^2\Phi}{dr^{*2}}+\Big[\omega^2-V(r^*)\Big]\Phi=0,
\end{eqnarray}
where the potential $V(r)$ in $r$ coordinate is given by
\begin{eqnarray}\label{pote}
V(r)=\frac{7+4\ell^2m^2}{4\ell^8}r^6
+\frac{4k^2\ell^2-10\ell^2M-4\ell^4m^2M}{4\ell^8}r^4
-\frac{4k^2\ell^4M-3\ell^4M^2}{4\ell^8}r^2.
\end{eqnarray}
%%%%%%%%%%%%%%%%%%%%%%%%%%%%%%%%%%%%%%%%%%%%%%%%%%%%%%%%%%%%%%%%%%%%%%%%%
%%%%%%%%%%%%%%%%%%%%%%%%%%%%%%%%%%%%%%%%%%%%%%%%%%%%%%%%%%%%%%%%%%%%%%%%%
\begin{figure*}[t!]
   \centering
   \includegraphics{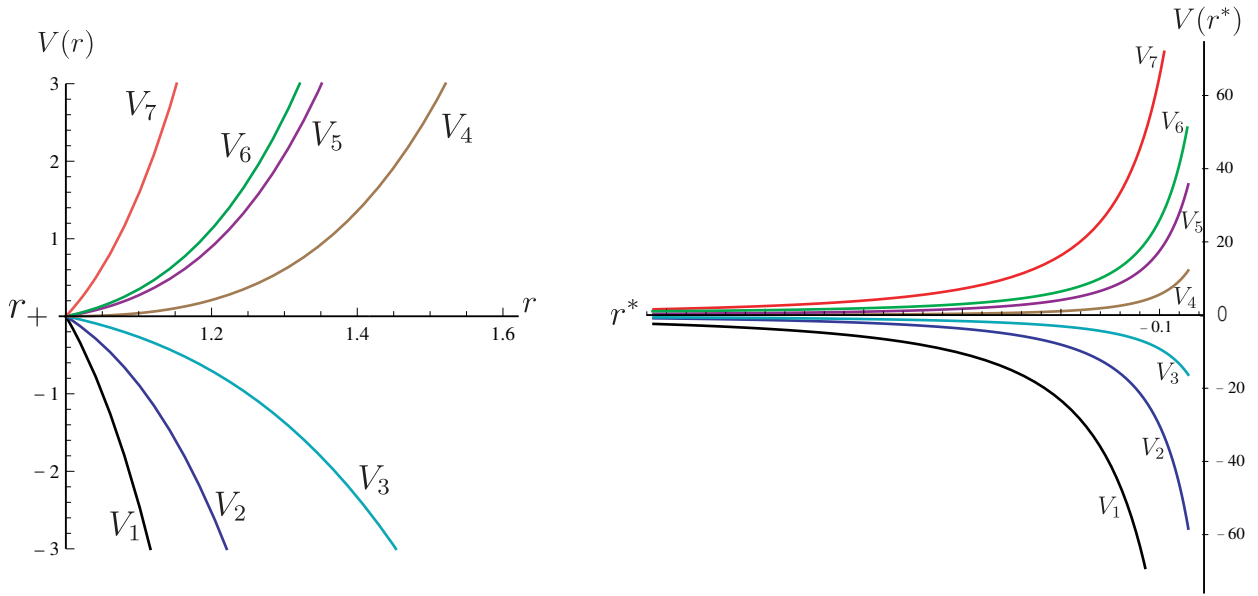}
\caption{Potential $V$ as a function of $r$ (left panel) and $r^*$
(right panel) with $M=1$, $\ell=1$, and $k=0$. In these figures,
$V_{1\sim7}$ correspond to $m^2\ell^2=-9,-4,-2,-1,-1/2,0,2$,
respectively. For $ m^2\ell^2\ge -1$, the  potentials ($V_{4\sim7}$)
are always positive for the whole range of $r_+\le r<\infty$ or
$-\infty< r^*\le0$. However, the stability is extended to
$V_2(m^2\ell^2=-4)$ because the scalar field is propagating in the
Lifshitz spacetimes.}
\end{figure*}
%%%%%%%%%%%%%%%%%%%%%%%%%%%%%%%%%%%%%%%%%%%%%%%%%%%%%%%%%%%%%%%%%%%%%%%%%
%%%%%%%%%%%%%%%%%%%%%%%%%%%%%%%%%%%%%%%%%%%%%%%%%%%%%%%%%%%%%%%%%%%%%%%%%
We could not obtain an analytic expression  $V(r^*)$ written by
$r^*$ from (\ref{pote}) because it is difficult to express $r$ in
terms of $r^*$ as  (\ref{toto}) does show.  However, $V(r^*)$ can be
plotted in a parametric way by using  (\ref{toto}) and (\ref{pote}).
Fig. 1 shows that for $ m^2\ell^2\ge -1$, the potential is always
positive, which  implies that the Lifshitz black hole is obviously
stable under the scalar perturbation. Actually, there is
correspondence between $V(r)$ and $V(r_*)$ for each $m^2$ [color
matching], implying that the stability criterion remains unchanged.
If one considers a scalar propagation in flat spacetimes, the
stability condition is just the non-tachyonic mass of $m^2 \ge 0$.
However, the stability condition was extended to the
Breitenlohner-Freedman (BF) bound ($m^2 \ge m^2_{\rm BF}=-1/\ell^2$)
in AdS$_3$ spacetimes~\cite{Breitenlohner:1982bm}, even if its
potential is negative in AdS$_3$ spacetimes. Similarly, as was first
mentioned in~\cite{L-1}, the BF bound can be extended to the
Lifshitz bound \be \label{Lib} m^2 \ge m^2_{\rm
Lif}=-\frac{4}{\ell^4} \ee in Lifshitz
spacetimes~\cite{Abdalla:2011fd}. In other words, the Lifshitz black
hole is unstable against the scalar propagation with mass ($m^2<
-4/\ell^2$) which is considered as the tachyonic instability.

In order  to understand the Lifshitz bound (\ref{Lib}) more
explicitly, we investigate  the potential (\ref{pote}) at large $r$
limit ($r_*\to 0$) which  is approximated as an inverse square
potential
\begin{eqnarray}\label{pote1}
V(r_*)\sim \frac{\xi}{r_*^2}
\end{eqnarray}
with \be \xi=\frac{7+4m^2\ell^2}{36}. \ee This is because the first
term in (\ref{pote}) dominates in this limit.   Then, Eq.
(\ref{sch}) becomes the Schr\"{o}dinger equation with the inverse
square potential.  It is known that for this type of Schr\"{o}dinger
equation,  the stability of the scalar field is determined  by the
condition of  $\xi\ge-1/4$ \cite{Case}  which yields the Lifshitz
bound~\footnote{Similarly, in the AdS$_{3}$ spacetimes \cite{Moroz},
a permitted range for the massive scalar field with the inverse
square potential is given by the BF bound: $\xi\ge-1/4 \to m^2 \ge
m^2_{\rm BF}=-1/\ell^2.$}.

Finally, it is important to note that  the Lifshitz bound
corresponds to the stability condition obtained when applying the
quasinormal mode approach.  To see this explicitly, we consider the
full form of QNFs \cite{COP,MM}
\begin{eqnarray}
\omega&=&-i4\pi T_H\Bigg[-1-2n
-(4+m^2\ell^2)^{1/2}\nonumber\\&&\hspace*{5em}+\Big(7+\frac{3m^2\ell^2}{2}
+\frac{k^2}{2M}+(3+6n)(4+m^2\ell^2)^{1/2}+6n(n+1)\Big)^{1/2}\Bigg]
\label{quasi}
\end{eqnarray}
with $n=0,1,2,\cdots$. One  checks that for $s$-mode ($k=0$), the
quasinormal frequency $\omega$ becomes negative imaginary if
$m^2\ell^2$ satisfies the relation
\begin{eqnarray}
\Big[7+\frac{3m^2\ell^2}{2}
+(3+6n)(4+m^2\ell^2)^{1/2}+6n(n+1)\Big]^{1/2}&>&1+2n+(4+m^2\ell^2)^{1/2}\nonumber\\
&&\hspace*{-13em}\longrightarrow~~~ m^2\ell^2 \ge -4
\end{eqnarray}
which is exactly (\ref{Lib}).

\section{Absorption cross section}
In order to find  the absorption cross section, we introduce  a new
coordinate $x=(r^2-r_+^2)/r^2$, which is useful to find a solution
to Eq.(\ref{omain}).  Then,  the radial equation takes the form
\begin{eqnarray}\label{3Dre}
\rho''(x)+\frac{1}{x(1-x)}\rho'(x)+
\frac{\ell^2(\omega^2(1-x)^3-m^2M^3x)-k^2M^2(1-x)}
{4M^3(1-x)^2x^2}\rho(x)=0.
\end{eqnarray}
Here the prime (${}^{\prime}$) denotes the differentiation with
respect to the variable $x$. We note  that $r\in[r_+,\infty)$ is
mapped to a compact region of $ x\in [0,1)$. The solution to this
equation is given by the confluent Heun (HeunC) functions as
\begin{eqnarray}\label{mainH}
\rho(x)&=&C_1 x^{\alpha}(1-x)^{\beta} {\rm
HeunC}\Big[0,~2\alpha,~2(\beta-1),~\alpha^2,~
(\beta-1)^2+\frac{1}{4M^3}(k^2M^2+\omega^2\ell^2);~x\Big]
\nonumber \\
&&\hspace*{-4em} +~C_2 x^{-\alpha}(1-x)^{\beta} {\rm
HeunC}\Big[0,-2\alpha,~2(\beta-1),~\alpha^2,~
(\beta-1)^2+\frac{1}{4M^3}(k^2M^2+\omega^2\ell^2);~x\Big],
\label{1sol11}
\end{eqnarray} where $C_{1,2}$ are arbitrary
constants and
\begin{eqnarray}\label{abeta}
\alpha=i\frac{\omega \ell}{2M^{3/2}}=i\frac{\omega}{4\pi
T_H},~~\beta=1+\sqrt{1+\frac{m^2\ell^2}{4}}
\end{eqnarray}
with $T_H$ the Hawking temperature of Lifshitz black hole. In the
neighborhood of the horizon ($x=0$), using the property of ${\rm
HeunC}[0,~a,~b,~c,~d;~0]=1$ leads to $\rho_0(x)=C_1 x^{\alpha}+C_2
x^{-\alpha}$ which yields
\begin{eqnarray}\label{solmain}
\varphi_0&=& C_1~e^{-i\omega t}x^{\alpha}+C_2~e^{-i\omega
t}x^{-\alpha} \nonumber \\
&\sim&C_1~e^{-i\omega[t-\frac{1}{4\pi T_H}\ln x]}
+C_2~e^{-i\omega[t+\frac{1}{4\pi T_H}\ln x]}.
\end{eqnarray}
Here, the former corresponds to outgoing mode
($\mid_{x=0}\longrightarrow$), while the latter to ingoing mode
($\mid_{x=0}\longleftarrow$). Now we consider a scattering process
where an ingoing mode comes from the spatial infinity and interacts
with the black hole. Then, it is partially reflected backwards as
outgoing mode to the infinity and the rest is absorbed into the
black hole horizon. One way to achieve this goal is to introduce  a
purely ingoing wave at the horizon and then, carefully to decompose
it into ingoing and outgoing waves in the large $r$ region. The
other approach  is that we begin with  ingoing and outgoing modes at
the horizon and a purely outgoing mode at the spatial infinity.  It
is known that two pictures provide the same absorbtion cross
section~\cite{Chen:2012zn}. In this work, we use the former method
to derive the absorption cross section of Lifshitz black hole.

To have  the ingoing mode at the horizon, the constant $C_1$ is set
to be zero in Eq.(\ref{solmain}).  We use the connection formula for
the HeunC functions~\cite{MM} to develop ingoing and outgoing modes
at infinity ($x\to1)$. Then, Eq.(\ref{mainH}) becomes
\begin{eqnarray}
&&\hspace*{-1.5em}\rho_1(x)=C_2\Bigg((1-x)^{\beta}
\frac{\Gamma(1-2\alpha)\Gamma(2-2\beta)}
{\Gamma(3-2\beta+K)\Gamma(-2\alpha-K)}
+(1-x)^{2-\beta}\frac{\Gamma(1-2\alpha)\Gamma(2\beta-2)}
{\Gamma(2\beta-1+S)\Gamma(-2\alpha-S)}\Bigg)\nonumber\\
\label{rhoeq1}&&
\end{eqnarray}
where $K$ and $S$ will be determined  as the solutions to two
algebraic equations
\begin{eqnarray}
K^2+(3+2\alpha-2\beta)K+2\alpha-2\beta+2-\epsilon+\frac{\alpha^2}{2}~=~0,
\label{Keq}\\
S^2+(2\alpha+2\beta-1)S-2\alpha+4\alpha\beta-\epsilon+\frac{\alpha^2}{2}
~=~0\label{Seq}
\end{eqnarray}
with \be
\epsilon=\frac{1}{2}[1-(1-2\alpha)(2\beta-1)]-(\beta-1)^2-\frac{1}{4M^3}(k^2M^2+\omega^2\ell^2).
\ee

Another way to find the solution at infinity is to start  with  the
equation  directly
\begin{eqnarray}\label{rhoeq}
\rho_\infty''(r)+\frac{5}{r}\rho_\infty'(r)+
\frac{\ell^2}{r^2}\left(\frac{\ell^6\omega^2}{r^6}-\frac{k^2}{r^2}-m^2\right)
\rho_\infty(r)=0,
\end{eqnarray}
which was obtained by setting $r_+=0$ in Eq.(\ref{omain}). In order
to find an analytic solution to (\ref{rhoeq}), one has to focus on
$k=0$ ($s$-mode). For $k\not=0$, it seems difficult to solve the
equation.  In this case,  the $s$-mode solution is given in terms of
Bessel functions
\begin{eqnarray}\label{rho-1}
\rho_\infty^{s}(r)=\left(\frac{\ell^4\omega}{2r^3}\right)^{\frac{2}{3}}
\Bigg[C_3J_{-\gamma}\left(\frac{\ell^4\omega}{3r^3}\right)
\Gamma(1-\gamma)+C_4J_{\gamma}\left(\frac{\ell^4\omega}{3r^3}\right)
\Gamma(1+\gamma)\Bigg],
\end{eqnarray}
where $C_{3,4}$ are integration constants and $\gamma$ is
\begin{eqnarray}
\gamma=\frac{2}{3}\sqrt{1+\frac{m^2\ell^2}{4}}=\frac{2}{3}(\beta-1).
\end{eqnarray}
For  large $r$, the solution (\ref{rho-1}) can be written as
\begin{eqnarray}\label{rho5}
\rho_\infty^{s}(r)=\tilde{C}_3\left(\frac{1}{r^2}\right)^{2-\beta}
+\tilde{C}_4\left(\frac{1}{r^2}\right)^{\beta}
\end{eqnarray}
with $\tilde{C}_{3,4}$ constants.

It is worth noting   that to have a regular behavior (normalizable
solution) at $r\to\infty$~\cite{Gonzalez:2010ht}, $\beta$
 should be confined  to $1\le \beta \le 2$
which corresponds to \be \label{sr} -\frac{4}{\ell^2}=m^2_{\rm Lif}
\le m^2 \le 0.\ee
 Imposing the  regularity at infinity restricts the stability
condition (\ref{Lib}) to a smaller region (\ref{sr}). According to
the AdS$_3$-analysis~\cite{Klebanov:1999tb}, it is shown in the
context of AdS/CFT correspondence that if the mass of a bulk scalar
lies in the interval \be -\frac{1}{\ell^2}=m^2_{\rm BF} \le m^2 \le
0, \ee a single gravity theory in the bulk describes two different
conformal field theories on the boundary. Similarly, we might
confine the allowed mass range to (\ref{sr}) in Lifshitz spacetimes.
In other words, (\ref{sr})  is necessary to compute the absorption
cross section of a massive scalar field propagating in the  Lifshitz
spacetimes.

We are now  in a position to obtain  the absorption cross section.
For this purpose, we have to decompose $\tilde{C}_{3,4}$ into the
ingoing and outgoing coefficients. However, distinguishing  between
ingoing and outgoing modes at asymptotic region is not a trivial
task because the spacetime is Lifshitz.  Following the ansatz in
asymptotically AdS spacetimes,  we introduce new constant parameters
$C_{{\rm in}}$ and $C_{{\rm out}}$ given by
\cite{Birmingham:1997rj,Gonzalez:2010ht}
\begin{eqnarray}\label{decom}
\tilde{C}_{3}=C_2\Big(C_{{\rm in}}+C_{{\rm
out}}\Big),~~\tilde{C}_{4}=icC_2\Big(C_{{\rm out}}-C_{{\rm in}}\Big),
\end{eqnarray}
where $c$ is a parameter with the length dimension $[L]^{4\beta-4}$.
For the ansatz (\ref{decom}),  comparing (\ref{rho5}) with
(\ref{rhoeq1}) leads to
\begin{eqnarray}
C_{{\rm in}}=\frac{\Gamma_{2}}{2}+i\frac{\Gamma_1}{2c},~~~C_{{\rm
out}}=\frac{\Gamma_{2}}{2}-i\frac{\Gamma_1}{2c},
\end{eqnarray}
where $\Gamma_{1,2}$ are given by
\begin{eqnarray}
\Gamma_1&=&r_+^{2\beta}\frac{\Gamma{(1-2\alpha)}\Gamma{(2-2\beta)}}
{\Gamma{(3-2\beta+K)}\Gamma{(-2\alpha-K)}},\label{ga1}\\
\Gamma_2&=&r_+^{4-2\beta}\frac{\Gamma{(1-2\alpha)}\Gamma{(2\beta-2)}}
{\Gamma{(2\beta-1+S)}\Gamma{(-2\alpha-S)}}\label{ga2}.
\end{eqnarray}
It is well-known that the conserved flux ${\cal F}(r)$ is defined by
\begin{eqnarray}
{\cal F}(r)=\frac{\sqrt{-g}g^{rr}}{2i}
\left(\rho^{*}\partial_{r}\rho-\rho\partial_{r}\rho^{*}\right).
\end{eqnarray}
Using this expression, the absorption coefficient (${\cal A}$) is
given by
\begin{eqnarray}
{\cal A}=\left|\frac{{\cal F}^{{\rm in}}_{r_+}}{{\cal F}^{{\rm
in}}_\infty} \right|,
\end{eqnarray}
where ${\cal F}_{r_+}$ and ${\cal F}_{\infty}$ denote the
flux at the horizon and at  asymptotic region $(r\to \infty)$,
respectively. These are computed as
\begin{eqnarray}
{\cal F}^{{\rm in}}_{r_+}=-\omega r_{+}|C_2|^2,~~~ {\cal
F}_{\infty}=\frac{4c(\beta-1)|C_2|^2}{\ell^4}\Big[|C_{\rm
in}|^2-|C_{\rm out}|^2\Big].
\end{eqnarray}
Then, we have ${\cal A}$ as
\begin{eqnarray}
{\cal A}=\frac{\omega \ell^4 r_{+}}{4c(\beta-1)|C_{\rm in}|^2}.
\end{eqnarray}
Finally,  the absorption cross section  $\sigma_{{\rm
abs}}$ takes the form
\begin{eqnarray}\label{abs}
\sigma_{{\rm abs}}=\frac{{\cal A}}{\omega}=\frac{ \ell^4
r_{+}}{4c(\beta-1)|C_{\rm in}|^2},
\end{eqnarray}
which looks like an unclear form.

Hence, we consider  the absorption cross section in the low energy
and massless limits. To investigate the absorption cross section in
the limits of $\omega\to0$ and $m\to0$ (equivalently, $ \alpha\to0$
and $\beta\to2)$, we rewrite Eq.(\ref{abs}) as
\begin{eqnarray} \label{abs2}
\sigma_{{\rm abs}}=\frac{ \ell^4
r_{+}}{c(\beta-1)|\Gamma_2+i\Gamma_1/c|^2},
\end{eqnarray}
where $\Gamma_{1,2}$ are given by (\ref{ga1}) and (\ref{ga2}). We
mention  that in the limits of  $\omega\to0$ and $~m\to0$,  the
parameters $K$ and $S$ are determined  as
\begin{eqnarray}\label{KS}
K=0~~{\rm or}~~1,~~~~~S=-1~{\rm or}~-2,
\end{eqnarray}
by solving (\ref{Keq}) and (\ref{Seq}). There are four combinations
of $(K,S)=(0,-1),(0,-2),(1,-1)$ and $(1,-2)$ which show a feature of
the Lifshitz black hole.   It turns out that four  combinations
provide a single combination for $\Gamma_1$ and $\Gamma_2$ as
\begin{eqnarray}
\Gamma_1 = 0,~~\Gamma_2= 1\label{ga-2}.
\end{eqnarray}
Substituting this into (\ref{abs2}) leads to  the absorption cross section
\begin{eqnarray}
\sigma^{\{\omega\to 0,~m\to 0\}}_{{\rm abs}}=2\pi r_+=A,
\end{eqnarray}
where we have chosen  $c= \frac{\ell^4}{2\pi}$ for recovering the
horizon area. We have to say that  there is no information to fix
$c$ in the wave equation approach.

Consequently,  it is shown  that the absorption cross section for
scattering of  the scalar  off  the $z=3$ Lifshitz black hole is
given by  the area of the horizon in the limits of $\omega\to0$ and
$~m\to0$.

\section{Discussions}
First of all, we would like to mention the stability of Lifshitz
black hole in three dimensions. Even though the potential is
positive definite for $m^2\ell^2 \ge -1$, the stability condition is
given by the Lifshitz bound (\ref{Lib})  when using a minimally
coupled scalar propagating in the Lifshitz black hole spacetimes.
This stability condition was confirmed by the quasinormal mode
approach to the Lifshitz black hole.

For the mass range (\ref{sr}), we have computed the absorption cross
section  by considering scattering of a minimally coupled scalar off
the Lifshitz black hole. The  absorption cross section reduces to
the horizon area in the low energy and massless limits of $s$-wave
propagation, indicating that the Lifshitz black hole also satisfies
the universality of low-energy absorption cross section for black
holes.

Finally, we propose that the decaying rate  is given
by~\cite{Birmingham:1997rj,LKM} \be \label{dr} \Gamma_{\rm
Lif}=\frac{\sigma_{\rm abs}}{e^{\frac{\omega}{T_H}}-1}, \ee where
$\sigma_{\rm abs}$ is given by (\ref{abs2}).  This decaying rate
could be calculated by the boundary field theory if the latter is
known~\cite{Gubser:1997cm}. However, we do not know exactly the
boundary field theory which may exist  by presuming the Lif/CFT
correspondence.  We expect to recover the proposed decaying rate
(\ref{dr}) from the boundary field theory approach in the near
future.

\section*{Acknowledgement}

 This work was supported by the National Research Foundation of Korea
(NRF) grant funded by the Korea government (MEST) through the Center
for Quantum Spacetime (CQUeST) of Sogang University with Grant
No.2005-0049409. Y. Myung  was partly supported by the National
Research Foundation of Korea (NRF) grant funded by the Korea
government (MEST) (No.2011-0027293).

\end{document}